\documentclass[prd,superscriptaddress,amsfonts,amssymb,amsmath,showpacs,twocolumn]{revtex4-2}
\usepackage{bm}
\usepackage{amsfonts}
\usepackage{latexsym}
\usepackage[latin1]{inputenc}
\usepackage{graphicx}
\usepackage{amsmath}
\usepackage{palatino}
\usepackage{mathpazo}
\usepackage{textcomp}
\usepackage{float}
\usepackage{booktabs}
\usepackage{dcolumn}
\usepackage{hyperref}
\hypersetup{colorlinks,citecolor=blue}
\usepackage{amsmath}
\usepackage{xcolor}
\usepackage{orcidlink}
\usepackage[caption=false]{subfig}
\usepackage{commath}
\usepackage{ragged2e}
\def\jnl@style{\it}
\def\aaref@jnl#1{{\jnl@style#1}}

\def\aaref@jnl#1{{\jnl@style#1}}

\def\aj{\aaref@jnl{AJ}}                   
\def\apj{\aaref@jnl{ApJ}}                 
\def\apjl{\aaref@jnl{ApJ}}                
\def\apjs{\aaref@jnl{ApJS}}               
\def\apss{\aaref@jnl{Ap\&SS}}             
\def\aap{\aaref@jnl{A\&A}}                
\def\aapr{\aaref@jnl{A\&A~Rev.}}          
\def\aaps{\aaref@jnl{A\&AS}}              
\def\mnras{\aaref@jnl{Mon.~Not.~Roy.~Astron.~Soc.}}             
\def\prd{\aaref@jnl{Phys.~Rev.~D}}        
\def\prc{\aaref@jnl{Phys.~Rev.~C}}  
\def\prl{\aaref@jnl{Phys.~Rev.~Lett.}}    
\def\qjras{\aaref@jnl{QJRAS}}             
\def\skytel{\aaref@jnl{S\&T}}             
\def\ssr{\aaref@jnl{Space~Sci.~Rev.}}     
\def\zap{\aaref@jnl{ZAp}}                 
\def\nat{\aaref@jnl{Nature}}              
\def\aplett{\aaref@jnl{Astrophys.~Lett.}} 
\def\apspr{\aaref@jnl{Astrophys.~Space~Phys.~Res.}} 
\def\physrep{\aaref@jnl{Phys.~Rep.}}      
\def\physscr{\aaref@jnl{Phys.~Scr}}       
\def\commat{\aaref@jnl{Comm.~Math.~Phys.}}              
\def\science{\aaref@jnl{Science}}               
\def\cqg{\aaref@jnl{Classical Quant.~Grav.}}            
\def\jpcs{\aaref@jnl{JPCS}}                                     
\def\ijmpd{\aaref@jnl{Int.~J.~Mod.~Phys.~D}}                    
\def\grg{\aaref@jnl{Gen.~Relat.~Gravit.}}               
\def\rpp{\aaref@jnl{Rep.~Prog.~Phys.}}          
\def\npa{\aaref@jnl{Nucl.~Phys.~A}}        
\def\lrr{\aaref@jnl{Living Rev.~Rel.}}                   
\def\jcap{\aaref@jnl{J.~Cosmology Astropart.~Phys.}}    
\def\rmp{\aaref@jnl{Rev.~Mod.~Phys.}}   
\def\epjc{\aaref@jnl{Eur.~Phys.~J.~C}}

\allowdisplaybreaks[1]

\begin{document}
\color{red}

\title{Statefinder Analysis of Symmetric Teleparallel Cosmology}

\author{Raja Solanki\orcidlink{0000-0001-8849-7688}}
\email{rajasolanki8268@gmail.com}
\affiliation{Department of Mathematics, Birla Institute of Technology and
Science-Pilani,\\ Hyderabad Campus, Hyderabad-500078, India.}
\author{P.K. Sahoo\orcidlink{0000-0003-2130-8832}}
\email{pksahoo@hyderabad.bits-pilani.ac.in}
\affiliation{Department of Mathematics, Birla Institute of Technology and
Science-Pilani,\\ Hyderabad Campus, Hyderabad-500078, India.}

\date{\today}
\begin{abstract}
Statefinder diagnostic is a convenient method that can differentiate between the various dark energy models. In this article, we analyze the statefinder parameters in symmetric teleparallel cosmology. The $f(Q)$ gravity theory is an alternative theory to GR, where gravitational interactions attribute to the non-metricity scalar $Q$. In the present work, we consider two $f(Q)$ models which contains a linear and a non-linear form of non-metricity scalar, specifically, $f(Q)=\alpha Q + \frac{\beta}{Q}$ and $f(Q)=\alpha Q + \beta Q^2$, where $\alpha$ and $\beta$ are free parameters and then statefinder parameters $(r,s)$ are evaluated. We plot the trajectory of our models in the $r-s$ plane. In addition, we analyze the physical behavior of different cosmological parameters such as density, deceleration, and the EoS parameters. Further, we use Om diagnostic to differentiate the behavior of both $f(Q)$ models. We found that both $f(Q)$ models predicts that the present universe is accelerating due to the dark energy component evolving due to non-metricity. Moreover, model I represents phantom type behavior while model II follows quintessence scenario.

\end{abstract}


\maketitle

\section{Introduction}\label{sec1}

In the last few decades, cosmology has faced a dramatic change as more and more evidence \cite{Riess,Perlmutter,M.T,K.A} support that the dark sector dominates almost all energy content of the universe. The availability of high precision cosmological observations such as Baryon Acoustic Oscillations \cite{D.J.,W.J.}, Large scale structure \cite{T.Koivisto,S.F.}, Cosmic Microwave Background Radiation \cite{R.R.,Z.Y.}, and the Wilkinson Microwave Anisotropy Probe experiment \cite{C.L.,D.N.} indicates that the present universe is accelerating and the dark energy occupies nearly $68.3$\% of the entire universe while dark matter and baryonic matter occupies nearly $26.8$ \% and $4.9$\% of the total energy content of the universe. The root cause that triggering this accelerated expansion attributed to that dark energy (DE). Nowadays, several cosmologists have been attracted towards investigating the fundamental nature of dark energy. 

In General Relativity (GR), the simplest candidate for DE is the Einstein's cosmological constant $\Lambda$ \cite{Peb} which can be regarded as a fluid with high negative pressure and constant energy density $\varepsilon=\frac{\Lambda}{8\pi G}$ and such a fluid is characterized by the equation of state $\omega_{\Lambda}=-1$. Also, it is found that cosmological constant $\Lambda$ suffered from two delicate issues known as fine tuning and the cosmic coincidence problem. In the quantum field theory, the theoretical value of the vacuum energy \cite{S.W.} is $123$ order of magnitude larger than its observed value which is $10^{-47} GeV^4 $ \cite{Riess,Perlmutter}. The absence of a response mechanism that sets a very small value of the cosmological constant is referred to as a fine tuning problem. Further, the coincidence problem is referred to as the observed coincidence between the densities of dark matter and the dark energy while it is so different during the beginning of expansion \cite{P.S.}.
 
To address the above cosmological issues, dynamical DE models were proposed. The most popular and widely studied class of time-dependent dark energy model is the quintessence model having EoS parameter $\omega >-1$. Ratra and Peebles introduced the first quintessence model \cite{RP}. The quintessence description of dark energy is much different from $\Lambda$CDM as it is time-varying while $\Lambda$ is always constant. Another interesting dynamical DE model is phantom models which are characterized by the equation of state $\omega < -1$ \cite{M.S.,M.S.-2}. The phantom model represents the growing dark energy models which cause a big rip in the universe \cite{S.N.,A.V.}. Although plenty of dynamical DE models such as k-essence \cite{T.C.,C.A.}, Chaplygin gas \cite{M.C.,A.Y.}, Chameleon \cite{J.K.}, tachyon \cite{T.P.}, etc have been introduced in the literature.

On the other hand, like so many dark energy models started appearing, either qualitative or quantitative discrimination between the various dark energy models becomes necessary.  To address this issue, new geometrical diagnostic parameters proposed by V. Sahni et al. are known as statefinder parameters $(r,s)$ \cite{V.S.}. The statefinder parameters are defined as
\begin{equation}\label{1a}
r=\frac{\dddot{a}}{aH^3}
\end{equation}
and 
\begin{equation}\label{1b}
s=\frac{(r-1)}{3(q-\frac{1}{2})}
\end{equation}

where $q=-a\ddot{a} / \dot{a}^2$ is the deceleration parameter.

The well-known geometrical parameters namely the Hubble parameter and the deceleration parameter is generated from the cosmic scale factor and its first and second order derivative respectively. Therefore the parameter $r$ is the natural succeeding step beyond the deceleration parameter since it is generated from the scale factor and its third order derivative, while $s$ is the simple combination of $r$ and $q$. Statefinder analysis provides a useful graphical classification for current alternative models for gravity at cosmological scales. They are widely used to classify dark energy models in
different scenarios including theories with torsion \cite{Muj}, Einstein-Aether theory as a lorentz breaking theory \cite{Ant}, and even from holographic point of the view \cite{DM}. To get in touch with statefinder analysis, one can check References \cite{Chao,JX}. For the $\Lambda$CDM case, it is found that $(r,s)=(1,0)$.  Consider the sufficiently large times, when cosmological constant dominates the universe and contribution from matter energy become negligible, then the scale factor becomes $a(t)=exp(t)$ so that one can have $r=1$ and $s=0$. The deviation of the evolutionary trajectories of a given model in the $r-s$ plane from the $\Lambda$CDM point $(1,0)$ defines the distance of that model from $\Lambda$CDM one. Thus, for a given model, the statefinder pair $(r,s)$ can be calculated and the trajectory of that model can be drawn in the $r-s$ plane, and the difference in the trajectory from that of the $\Lambda$CDM gives the required discrimination. Furthermore, for different dark energy models, the different evolutionary trajectories in the $r-s$ plane have been found \cite{M.J.}. \\

Another alternative approach to address the late-time acceleration issue and to describe the origin of dark energy is to modify the action of GR, so called Modified theories of gravity. Till now several modified theories have been proposed. Some modified theories which have extensively investigated are $f(R)$ \cite{L.A.,SA,S.N.-2}, $f(T)$ \cite{R.F.,E.V.,X.R.}, $f(G)$ \cite{S.Nojiri,GC}, $f(R,\mathcal{T})$ \cite{Mor}, $f(R,G)$ \cite{AD} theories, etc, with $R$, $T$, $G$ and $\mathcal{T}$ being the Ricci, Torsion, Gauss-Bonnet and energy momentum scalars respectively.
In this article, we will work with the recently proposed $f(Q)$ theory of gravity with $Q$ being the non-metricity scalar \cite{J.B.}.

One can describe the gravitational interactions in the space-time manifold using three kinds of geometrical objects namely curvature, torsion, and non-metricity. In General Relativity (GR), gravitational interactions are attributed to space-time curvature. Another two possibilities torsion and non-metricity give the equivalent description of GR and the corresponding gravity are called Teleparallel and Symmetric Teleparallel Equivalent of GR. The $f(R)$ theory of gravity is a generalization of curvature-based gravity (GR) with vanishing torsion and non-metricity \cite{A.A.}. Similarly, the $f(T)$ gravity theory is a generalization of torsion-based gravity (the teleparallel equivalent of GR) with vanishing non-metricity and curvature \cite{GB}. Lastly, the $f(Q)$ gravity theory is a generalization of the symmetric teleparallel equivalent of GR with vanishing torsion and curvature \cite{Har}. Nowadays, investigations on the $f(Q)$ gravity theory frequently appear in the literature. Recently, Hassan et al. have discussed the traversable wormhole in symmetric teleparallel gravity \cite{Z}. Mandal et al. have analyzed the energy conditions and cosmography in $f(Q)$ gravity \cite{MS,MS-2}.
 
In this work, we are going to present a complete analysis of the statefinder diagnostic for symmetric teleparallel cosmology. This manuscript is organized as follows. In Sec \ref{sec2}, we present the motions equations in $F(Q)$ gravity. In Sec \ref{sec3}, statefinder parameters are evaluated for $F(Q)$ cosmology. In Sec \ref{sec4}, we consider two $F(Q)$ cosmological model and then statefinder parameters are evaluated. Further, a detailed physical analysis of cosmological parameters is done in this section. In Sec \ref{sec5},  we use Om diagnostic to differentiate the behavior of both $F(Q)$ models. Finally, we discuss our conclusions in Sec \ref{sec6}.

\section{Field Equation in $F(Q)$ gravity}\label{sec2}

The universe in $F(Q)$ gravity is described by following action:
\begin{equation}\label{2a}
S= \int{\frac{1}{2}F(Q)\sqrt{-g}d^4x} + \int{L_m\sqrt{-g}d^4x}
\end{equation}
where $F(Q)$ is an arbitrary function of the nonmetricity scalar $Q$,  $g$ is the determinant of the metric tensor $g_{\mu\nu}$ and $L_m$ is the Lagrangian density of matter.

Another key component to explain the symmetric teleparallel gravity is the non-metricity tensor, which is given as
\begin{equation}\label{2b}
Q_{\lambda\mu\nu} = \nabla_\lambda g_{\mu\nu}
\end{equation} 
and its two traces are shown below
\begin{equation}\label{2c}
Q_\alpha = Q_\alpha\:^\mu\:_\mu \: and\:  \tilde{Q}_\alpha = Q^\mu\:_{\alpha\mu}
\end{equation}
In addition, the superpotential (non-metricity conjugate) tensor is given by
\begin{equation}\label{2d}
4P^\lambda\:_{\mu\nu} = -Q^\lambda\:_{\mu\nu} + 2Q_{(\mu}\:^\lambda\:_{\nu)} + (Q^\lambda - \tilde{Q}^\lambda) g_{\mu\nu} - \delta^\lambda_{(\mu}Q_{\nu)}
\end{equation}
Then the trace of non-metricity tensor can be acquired as
\begin{equation}\label{2e}
Q = -Q_{\lambda\mu\nu}P^{\lambda\mu\nu} 
\end{equation}
Furthermore, the definition of the stress-energy tensor for the matter is 
\begin{equation}\label{2f}
T_{\mu\nu} = \frac{-2}{\sqrt{-g}} \frac{\delta(\sqrt{-g}L_m)}{\delta g^{\mu\nu}}
\end{equation}
For notational simpliciy, we define $ F_Q = \frac{dF}{dQ} $

The gravitational field equation obtained by varying the action \eqref{2a} with respect to the metric is given below
\begin{widetext}
\begin{equation}\label{2g}
\frac{2}{\sqrt{-g}}\nabla_\lambda (\sqrt{-g}F_Q P^\lambda\:_{\mu\nu}) + \frac{1}{2}g_{\mu\nu}F+F_Q(P_{\mu\lambda\beta}Q_\nu\:^{\lambda\beta} - 2Q_{\lambda\beta\mu}P^{\lambda\beta}\:_\nu) = -T_{\mu\nu}
\end{equation}
\end{widetext}

\section{State-Finder Parameters for $F(Q)$ Cosmology}\label{sec3}

Now, by assuming the cosmological principle, we describe our universe by the spatially isotropic and homogeneous flat FLRW metric \cite{Ryden}:
\begin{equation}\label{3a}
ds^2= -dt^2 + a^2(t)[dx^2+dy^2+dz^2]
\end{equation}
Here, $ a(t) $ is the scale factor that measures the cosmic expansion. One can find the non-metricity scalar by taking the trace of non-metricity tensor \eqref{2b} with respect to line element given by \eqref{3a} as
\begin{equation}\label{3b}
 Q= 6H^2  
\end{equation}

The stress-energy tensor for a perfect fluid distribution with respect to the metric \eqref{3a} is 

\begin{equation}\label{3c}
\mathcal{T}_{\mu\nu}=(\rho+p)u_\mu u_\nu + pg_{\mu\nu}
\end{equation}

Here, $\rho$ is the matter-energy density, $p$ is the usual pressure, and $u^\mu=(1,0,0,0)$ are components of the four velocity.

Then the Friedmann equations governing the dynamics of the universe are 

\begin{equation}\label{3d}
3H^2= \frac{1}{2F_Q}\left(-\rho + \frac{F}{2}\right)
\end{equation}
and
\begin{equation}\label{3e}
\dot{H}+3H^2+\frac{\dot{F_Q}}{F_Q}H = \frac{1}{2F_Q} \left(p+\frac{F}{2}\right)
\end{equation}  

We can rewrite the Friedmann equations for a non-relativistic matter dominated universe with the functional form  $F(Q)=-Q+f(Q)$, in a more suitable form as

\begin{equation}\label{3f}
H^2=\frac{1}{3} \left[ \rho_m+\rho_Q \right]
\end{equation}

\begin{equation}\label{3g}
\dot{H}=-\frac{1}{2} \left[ \rho_m+\rho_Q+p_Q \right]
\end{equation}

where 

\begin{equation}\label{3h}
\rho_Q=-\frac{f}{2}+Qf_Q
\end{equation}
and
\begin{equation}\label{3i}
p_Q=\frac{f}{2}-(2\dot{H}+Q)f_Q -2\dot{f_Q}H
\end{equation}

These equations \eqref{3f} and \eqref{3g} can be interpreted as the effective Friedmann equations along with an exotic fluid part coming from the non-metricity scalar, where $\rho_Q$ and $p_Q$ represents the energy density and pressure of the exotic fluid due to non-metricity.

Now the dimensionless density parameters corresponding to the matter and non-metricity density is defined by

\begin{equation}\label{3j}
  \Omega_m=\frac{\rho}{3H^2}, \ \ \ \Omega_Q=\frac{\rho_Q}{3H^2}, \ \  \   \Omega_m+\Omega_Q=1
\end{equation}

Then we can derive the expressions for the equation of state parameter and the deceleration parameters as 

\begin{equation}\label{3k}
\omega= \frac{p_Q}{\rho_Q} = -1 + \phi
\end{equation}
where 
\begin{equation}\label{3l}
\phi= 4\dot{H} \left( \frac{f_Q+2Qf_{QQ}}{f-2Qf_Q} \right)
\end{equation}
and
\begin{equation}\label{3m}
q=\frac{1}{2} (1+3\omega \Omega_Q)
\end{equation}

The statefinder parameter $r$ in terms of deceleration parameter can be written as

\begin{equation}\label{3n}
r=2q^2+q-\frac{\dot{q}}{H}
\end{equation}

Using equations \eqref{3k}-\eqref{3n} , we obtain the statefinder parameters for $F(Q)$ cosmology

\begin{widetext}
\begin{equation}\label{3o}
r=\frac{1}{2} \left[ \left\lbrace 1+3 \left(-1+\phi \right)\Omega_Q \right\rbrace ^2 +  \left\lbrace 1+3 \left(-1+\phi \right)\Omega_Q \right\rbrace \right]- \frac{3}{2H} \left[ \dot{\omega}\Omega_Q+ \left( -1+\phi \right) \dot{\Omega_Q} \right]
\end{equation}
\end{widetext}
and
\begin{equation}\label{3p}
s= \frac{2(r-1)}{9(-1+\phi)\Omega_Q}
\end{equation}

Furthermore, $F(Q)=-Q$ is the GR limit i.e, one can retrieve the usual Friedmann equations of GR for this functional form and the energy density and pressure due to non-metricity becomes zero, which is expected.

\section{Cosmological $F(Q)$ Models}\label{sec4}
\justify

In this section, we are going to analyze our results obtained in the previous section for some $f(Q)$ models.\\

\textbf{Model I:} We consider the following $f(Q)$ function which contains a linear and a non-linear form of non-metricity, specifically \cite{Harko}
 
\begin{equation}\label{4a}
f(Q)= \alpha Q + \frac{\beta}{Q}
\end{equation} 
where $\alpha$ and $\beta$ are free parameters.\\

Then for this specific choice of the function, one can obtain the following differential equation

\begin{equation}\label{4b}
\dot{H} \left( \alpha-1+\frac{\beta}{12H^4} \right) + \frac{3}{2} H^2  \left( \alpha-1-\frac{\beta}{12H^4} \right)=0
\end{equation}

Now, one can rewrite the Friedmann equations \eqref{3d} and \eqref{3e} in terms of dimensionless matter density parameter as

\begin{equation}\label{4c}
H^2=\frac{1}{12F_Q} \left[ -\Omega_m Q+ F \right]
\end{equation}

\begin{equation}\label{4d}
 \dot{H}= \frac{1}{4F_Q} \left[ \Omega_m Q-4F_{QQ} \dot{Q} H \right]
\end{equation}

The aim is to estimate the parameters of our proposed $f(Q)$ model that would be in agreement with the observed values of the cosmographic parameters. Then by using Friedmann equations \eqref{4c} and \eqref{4d} one can estimate

\begin{equation}\label{4e}
\alpha = 1-\frac{\Omega_{m0}}{2} \left[ 1+ \frac{3}{2 \left(1+q_0 \right)} \right]
\end{equation}
and
\begin{equation}\label{4f}
\beta=6\Omega_{m0} H_0^4 \left[ 1- \frac{3}{2 \left(1+q_0 \right)} \right]
\end{equation}

Now, we set $E(z)=\frac{H(z)}{H_0}$, then by using \eqref{4e} and \eqref{4f} in \eqref{4b}  we have the following differential equation in terms of dimensionless Hubble parameter

\begin{equation}\label{4g}
E' \left[ 1- \frac{1}{E^4} \frac{(2q_0-1)}{(2q_0+5)} \right] - \frac{3E}{2(1+z)}  \left[ 1+ \frac{1}{E^4} \frac{(2q_0-1)}{(2q_0+5)} \right] =0
\end{equation}

Now, by using equations \eqref{3j} and \eqref{3k} we have

\begin{equation}\label{4h}
\Omega_Q=  1-  \Omega_{m0}  \frac{(-1+2q_0+ E^4 (5+2q_0))}{4E^4(1+q_0)}
\end{equation}

\begin{widetext}
\begin{equation}\label{4i}
\omega= - \frac{8E^4 (2q_0^2+q_0-1)}{\{ 1-2q_0+E^4(5+2q_0)\} \{(2q_0-1) \Omega_{m0} + E^4 (5\Omega_{m0}-4+2q_0 (\Omega_{m0}-2)\}}
\end{equation}
\end{widetext}

The effective equation of state parameter for our model is given by

\begin{equation}\label{4j}
w_{eff}= \frac{p_{eff}}{\rho_{eff}}=\frac{p_Q}{\rho_m + \rho_Q}
\end{equation}

Here, $p_{eff}$ and $\rho_{eff}$ correspond to the total pressure and energy density of the universe. Then by using \eqref{3h} and \eqref{3i}, we have
 
\begin{equation}\label{4k}
w_{eff} =  \frac{4q_0-2}{1-2q_0+E^4 (5+2q_0)}
\end{equation}

By using the definition of deceleration parameter given in equation \eqref{3m}, we have

\begin{equation}\label{4l}
q=\frac{E^4 (5+2q_0) +10q_0 -5}{2E^4 (5+2q_0)-4q_0+2}
\end{equation}

Now, for our particular $f(Q)$ model we have the following expression of the statefinder parameters

\begin{widetext}
\begin{equation}\label{4m}
r=\frac{10 (1-2q_0)^2 + E^8 (5+2q_0)^2 + (4q_0^2+8q_0-5) \{ 7E^4 -12(1+z) E^3 E' \} }{\{ 1-2q_0 + E^4 (5+2q_0) \}^2}
\end{equation}
\end{widetext}

and

\begin{equation}\label{4n}
s=\frac{-3+3E^4 (5+2q_0) - 4(1+z)(5+2q_0)E^3E'}{3-6q_0+3E^4 (5+2q_0)}
\end{equation}

\justify \textbf{Numerical Results and Physical Aspects of Cosmological Parameters:}

We will solve the differential equation \eqref{4g} by the numerical algorithm. In the present work, we adopt the initial condition as $E(0)=1$, and we use $H_0=67.9 \: km/s/Mpc$ , $q_0=-0.5$, $\Omega_0=0.3$ \cite{Planck}. The following numerical plots are obtained using the above observed value of cosmographic parameters.

\begin{figure}[H]
\includegraphics[scale=0.62]{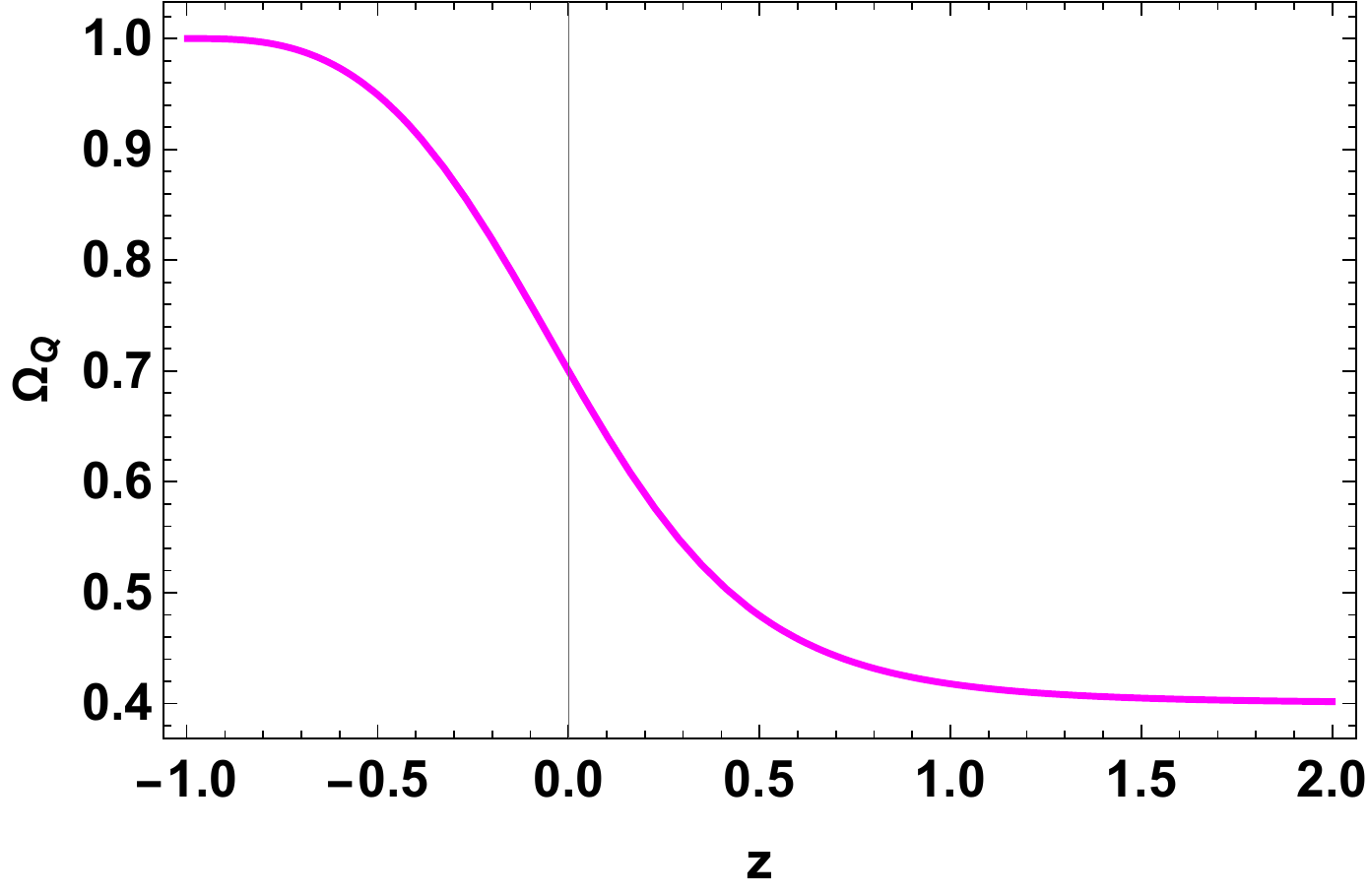}
\caption{Evolution profile of the cosmic density parameter vs redshift z.}\label{f1}
\end{figure}

\begin{figure}[H]
\includegraphics[scale=0.62]{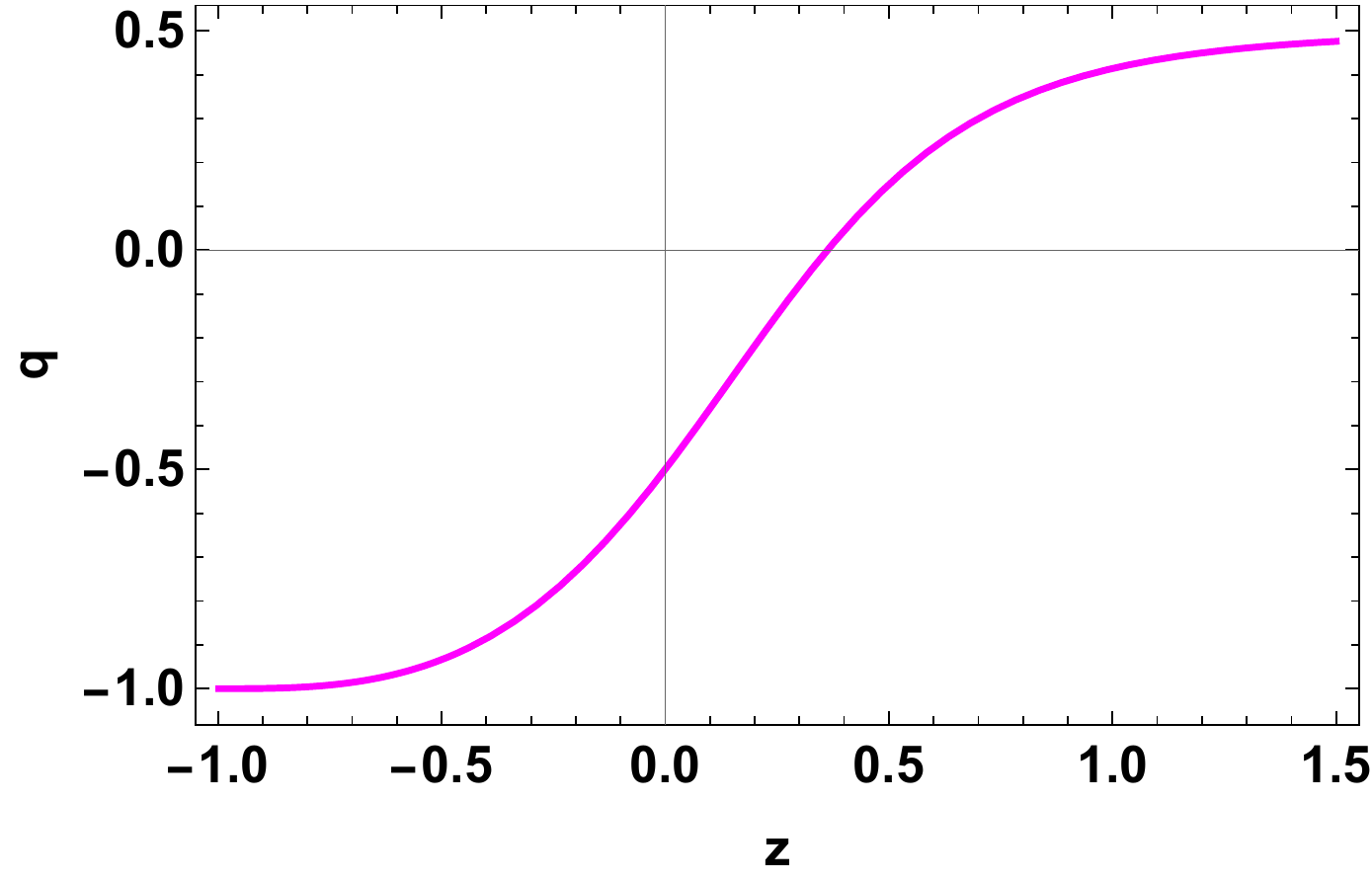}
\caption{Evolution profile of the deceleration parameter vs redshift z.}\label{f2}
\end{figure}

\begin{figure}[H]
\includegraphics[scale=0.65]{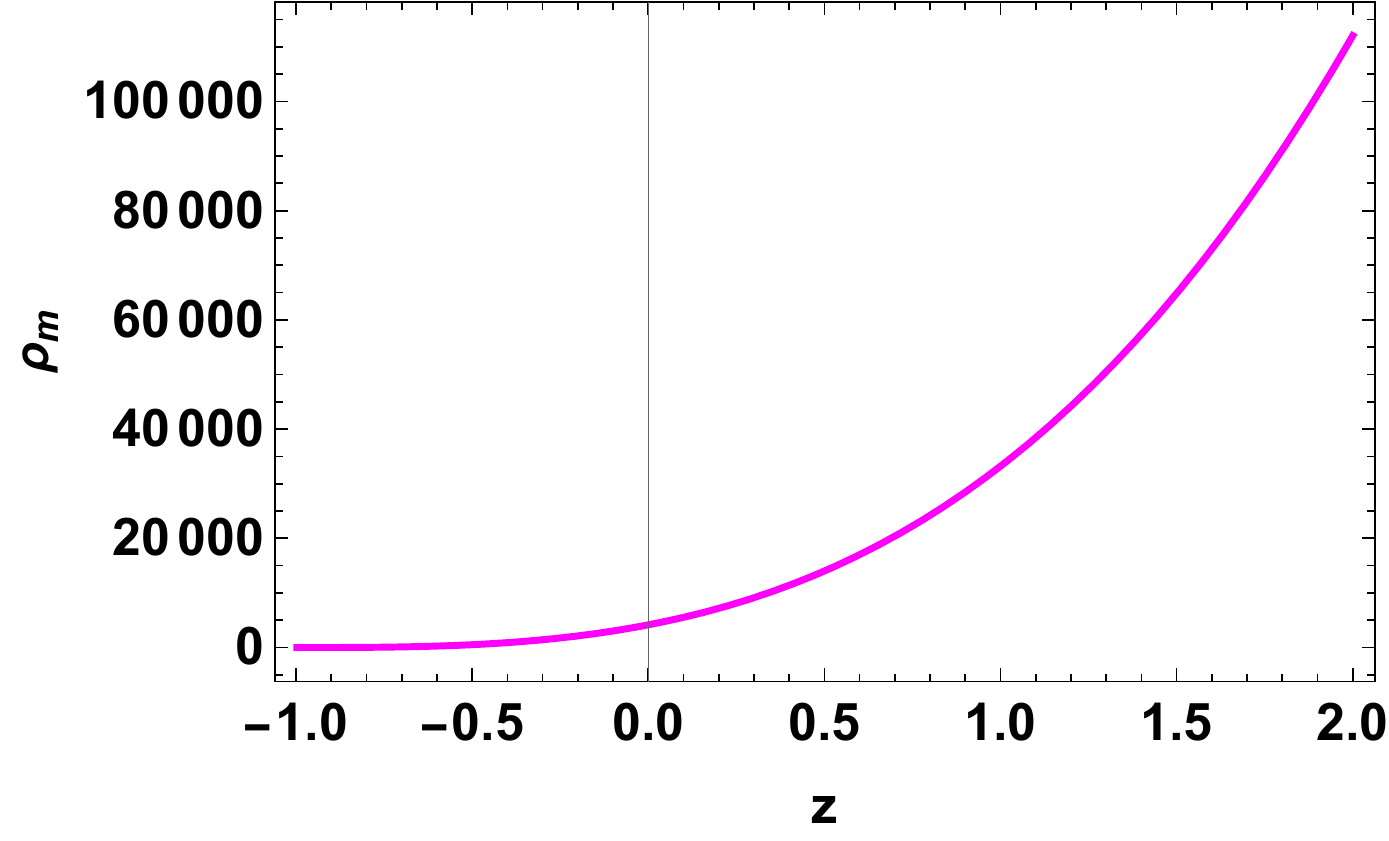}
\caption{Evolution profile of the cosmic energy density of matter vs redshift z.}\label{f3}
\end{figure}

\begin{figure}[H]
\includegraphics[scale=0.63]{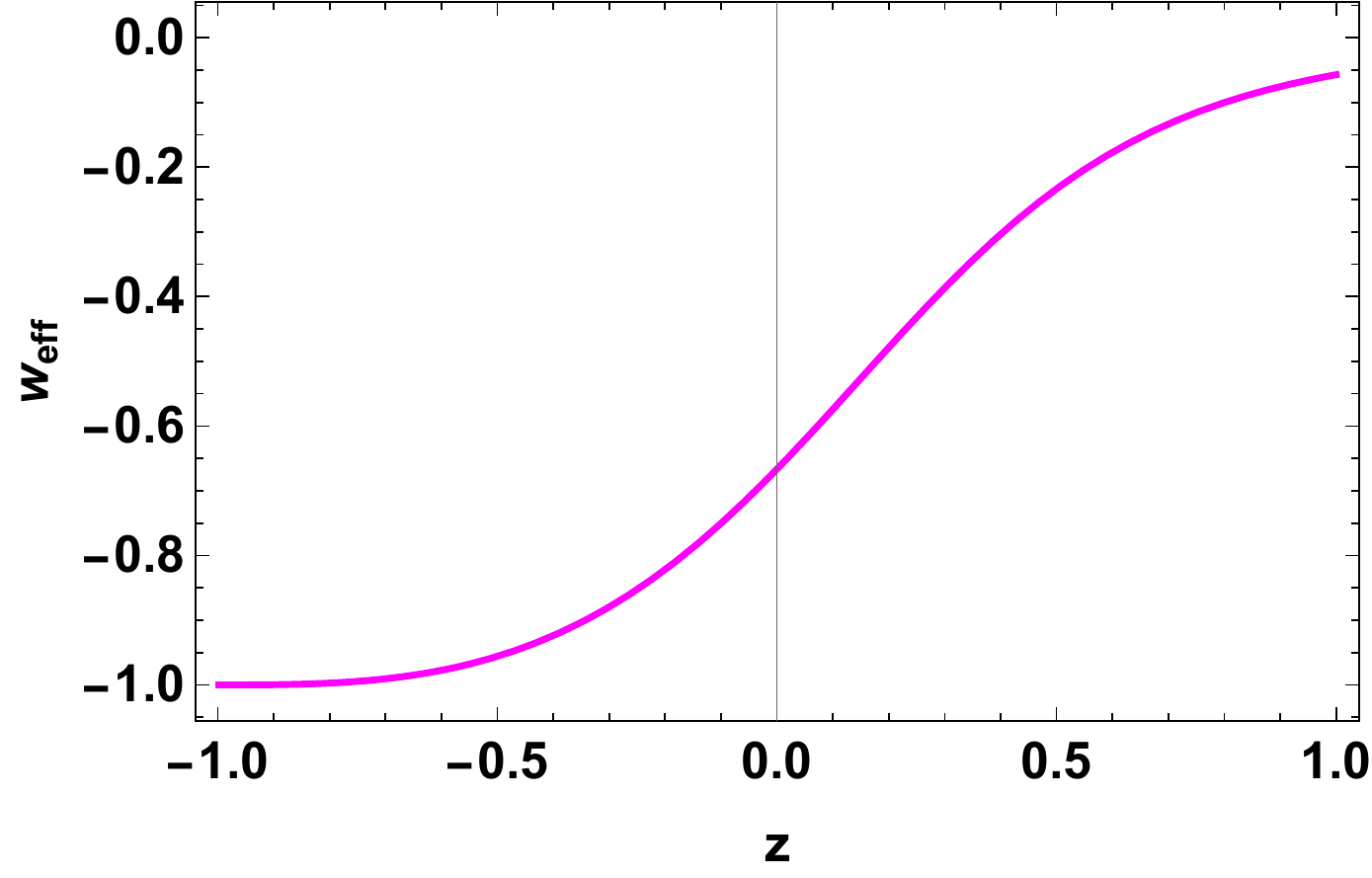}
\caption{Evolution profile of the effective EoS parameter vs redshift z.}\label{f4}
\end{figure}

\begin{figure}[H]
\includegraphics[scale=0.63]{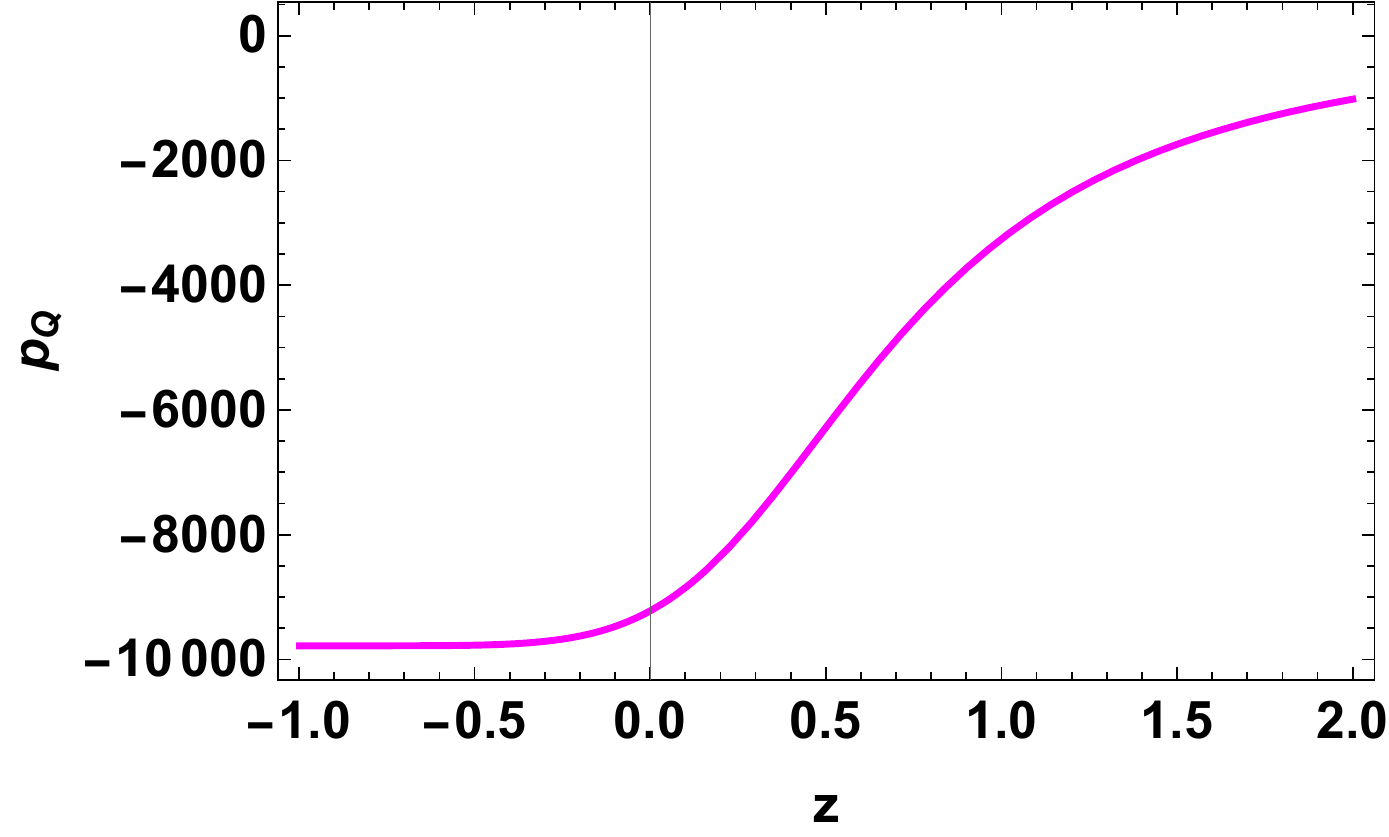}
\caption{Evolution profile of the cosmic pressure due to non-metricity vs redshift z.}\label{f5}
\end{figure}

\begin{figure}[H]
\includegraphics[scale=0.6]{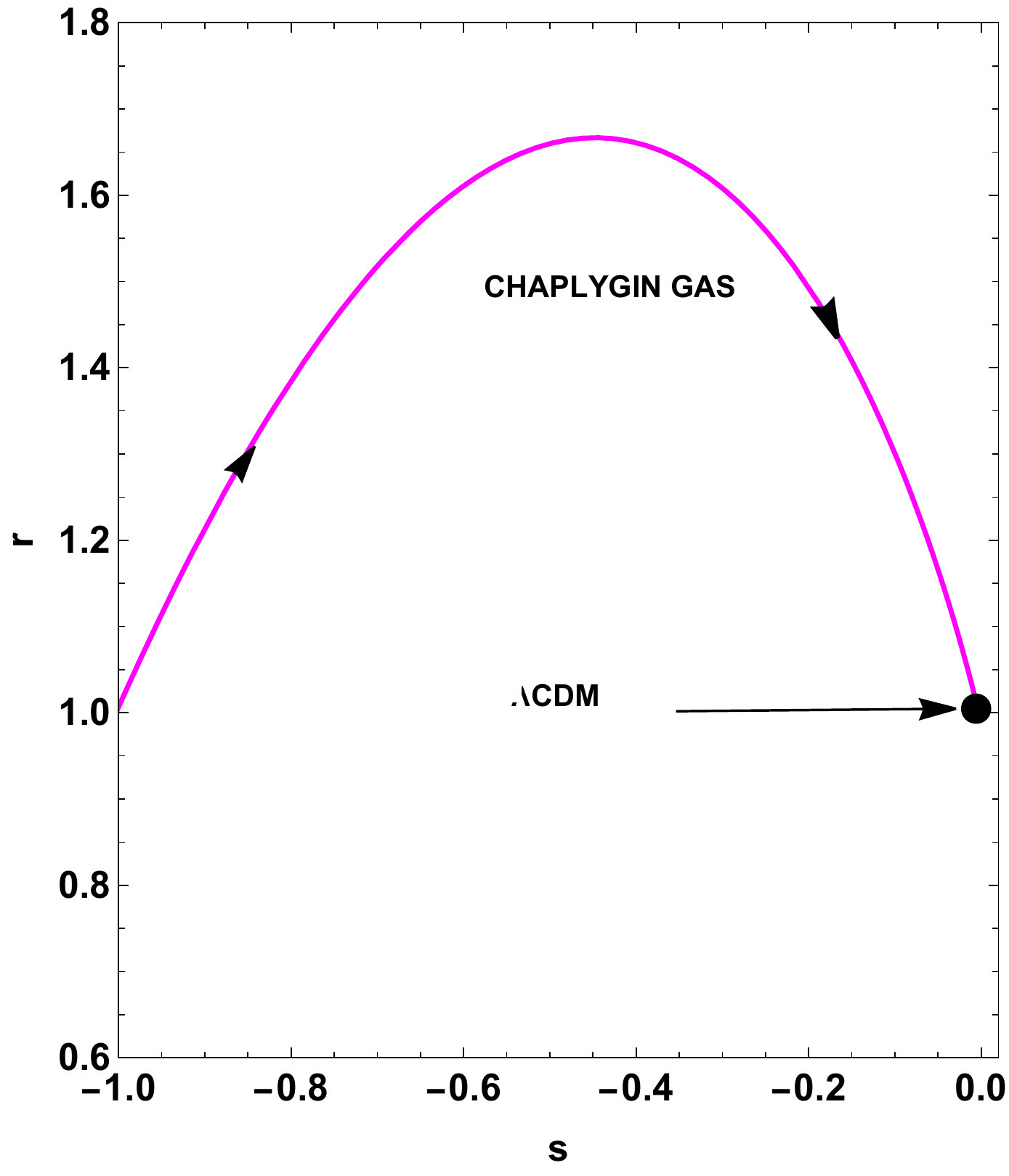}
\caption{Profile of the evolution trajectories of the given cosmological model in the $r-s$ plane . }\label{f6}
\end{figure}

From Fig \ref{f3} it is clear that the matter energy density of the universe decreases and vanishes in the far future with expansion of the universe while Fig \ref{f1} show that if the dark energy component is represented by the non-metricity then it will overcome all the energy content of the universe. Further, Fig \ref{f5} shows that the non-metricity component of the universe is of high negative pressure. The deceleration parameter is a key component to understand the expansion scenario of the universe. The plot for deceleration parameter in Fig \ref{f2} shows a phase transition from early deceleration to present acceleration of the universe. Further, from Fig \ref{f4} it is clear that the present universe is in the accelerating phase. 
The statefinder parameter describes the behavior of the dark energy component. A value $(r<1, s>0)$ represents quintessence type behavior of dark energy component, $(r=1, s=0)$ represents $\Lambda$CDM type behavior, and $(r>1, s<0)$ represents phantom type behavior. Fig \ref{f6} represents the evolution trajectories of given model which shows that our model lies in the chaplygin gas region $(r>1, s<0)$ and it will pass through the $\Lambda$CDM fixed point in the far future. \\

\textbf{Model II:}   We consider the following $f(Q)$ function which contains a linear and a non-linear form of non-metricity, specifically \cite{WKK}

\begin{equation}\label{5a}
f(Q)= \alpha Q + \beta Q^2
\end{equation} 
where $\alpha$ and $\beta$ are free parameters.

Then for this specific choice of the function, one can obtain the following differential equation

\begin{equation}\label{5b}
\dot{H} \left( \alpha -1+36 \beta H^2 \right) + \frac{3}{2} H^2  \left( \alpha -1+18 \beta H^2 \right)=0
\end{equation}

Again, by using Friedmann equations \eqref{4c} and \eqref{4d} one can estimate

\begin{equation}\label{5c}
\alpha = 1-\Omega_{m0} \left[ 2- \frac{3}{2 \left(1+q_0 \right)} \right]
\end{equation}
and
\begin{equation}\label{5d}
\beta=\frac{\Omega_{m0}}{18H_0^2} \left[ 1- \frac{3}{2 \left(1+q_0 \right)} \right]
\end{equation}

Then by using \eqref{5c} and \eqref{5d} in \eqref{5b} we have 

\begin{equation}\label{5e}
E' \left[ 1+2E^2 \frac{(1-2q_0)}{(1+4q_0)} \right] - \frac{3E}{2(1+z)}  \left[ 1+E^2 \frac{(1-2q_0)}{(1+4q_0)} \right] =0
\end{equation}

Now, by using equations \eqref{3j} and \eqref{3k} we have

\begin{equation}\label{5f}
\Omega_Q= \frac{2q_0 [1+\Omega_{m0} (E^2-2)]-\Omega_{m0} (E^2+1)+2}{2(1+q_0)}
\end{equation}

\begin{widetext}
\begin{equation}\label{5g}
\omega= - \frac{2(2q_0^2+q_0-1)E^2}{ \{E^2(4q_0-2)-4q_0-1\} \{2q_0 [ 1+ \Omega_{m0} (E^2-2) ]  - \Omega_{m0} (E^2+1) +2 \} }
\end{equation}
\end{widetext}

and the effective equation of state parameter for this model is given by
 
\begin{equation}\label{5h}
w_{eff} =  \frac{(1-2q_0)E^2}{ E^2(4q_0-2)-4q_0-1}
\end{equation}

By using the definition of deceleration parameter given in equation \eqref{3m}, we have

\begin{equation}\label{5i}
q=\frac{E^2 (1-2q_0) -4q_0 -1}{2E^2 (2q_0-1)-8q_0-2}
\end{equation}

The expression of the statefinder parameters for the given $f(Q)$ model are 

\begin{equation}\label{5j}
r=\frac{ (8q_0^2-2q_0-1) \{ E^2 +6(1+z) E E' \} }{ 2 \{ 2 E^2 (1-2q_0) +4q_0+1 \}^2}
\end{equation}

and

\begin{equation}\label{5k}
s=\frac{3E^3 (2q_0-1) -3E(1+4q_0)- 2(1+z)(1+4q_0)E'}{ 3E \{ 2E^2 (2q_0-1)-4q_0-1 \} }
\end{equation}

\justify \textbf{Numerical Results and Physical Aspects of Cosmological Parameters:}

Again, we will solve the differential equation \eqref{5e} by the numerical algorithm. We adopt the initial condition as $E(0)=1$, and we use $H_0=67.9 \: km/s/Mpc$ , $q_0=-0.5$, $\Omega_0=0.3$ \cite{Planck}. The following numerical plots are obtained using the above observed value of cosmographic parameters.

\begin{figure}[H]
\includegraphics[scale=0.62]{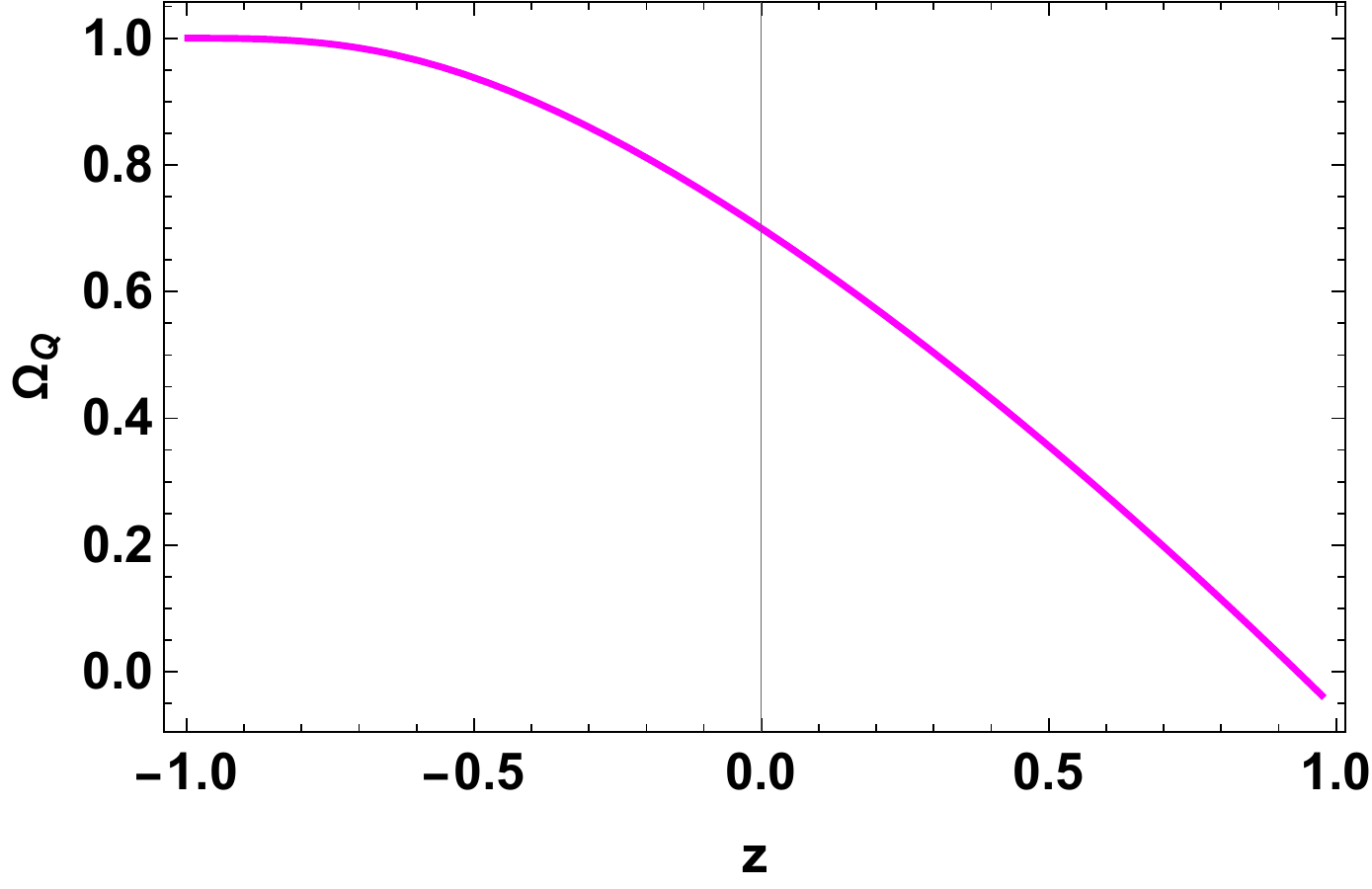}
\caption{Evolution profile of the cosmic density parameter vs redshift z.}\label{f8}
\end{figure}

\begin{figure}[H]
\includegraphics[scale=0.64]{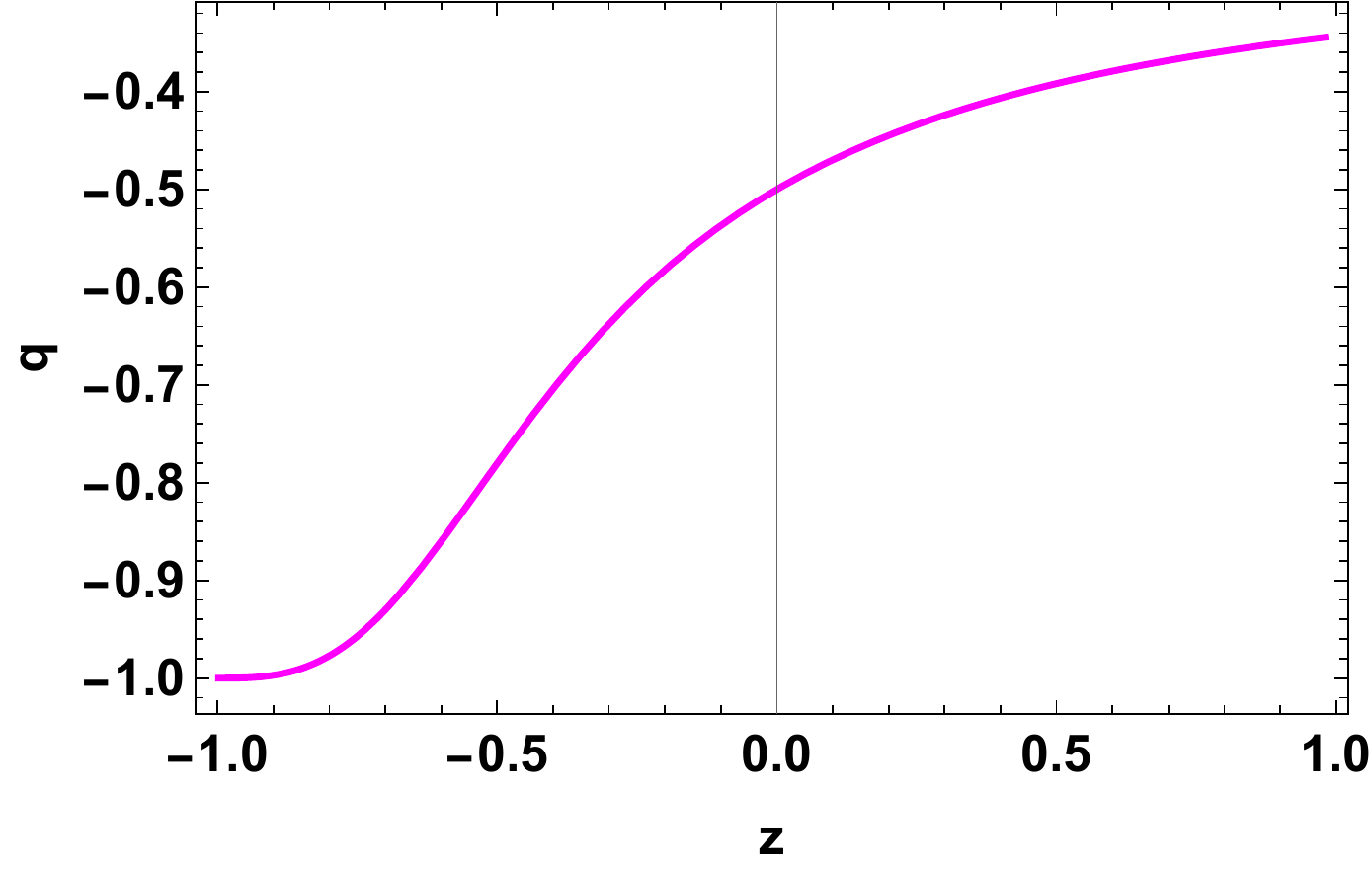}
\caption{Evolution profile of the deceleration parameter vs redshift z.}\label{f9}
\end{figure}

\begin{figure}[H]
\includegraphics[scale=0.65]{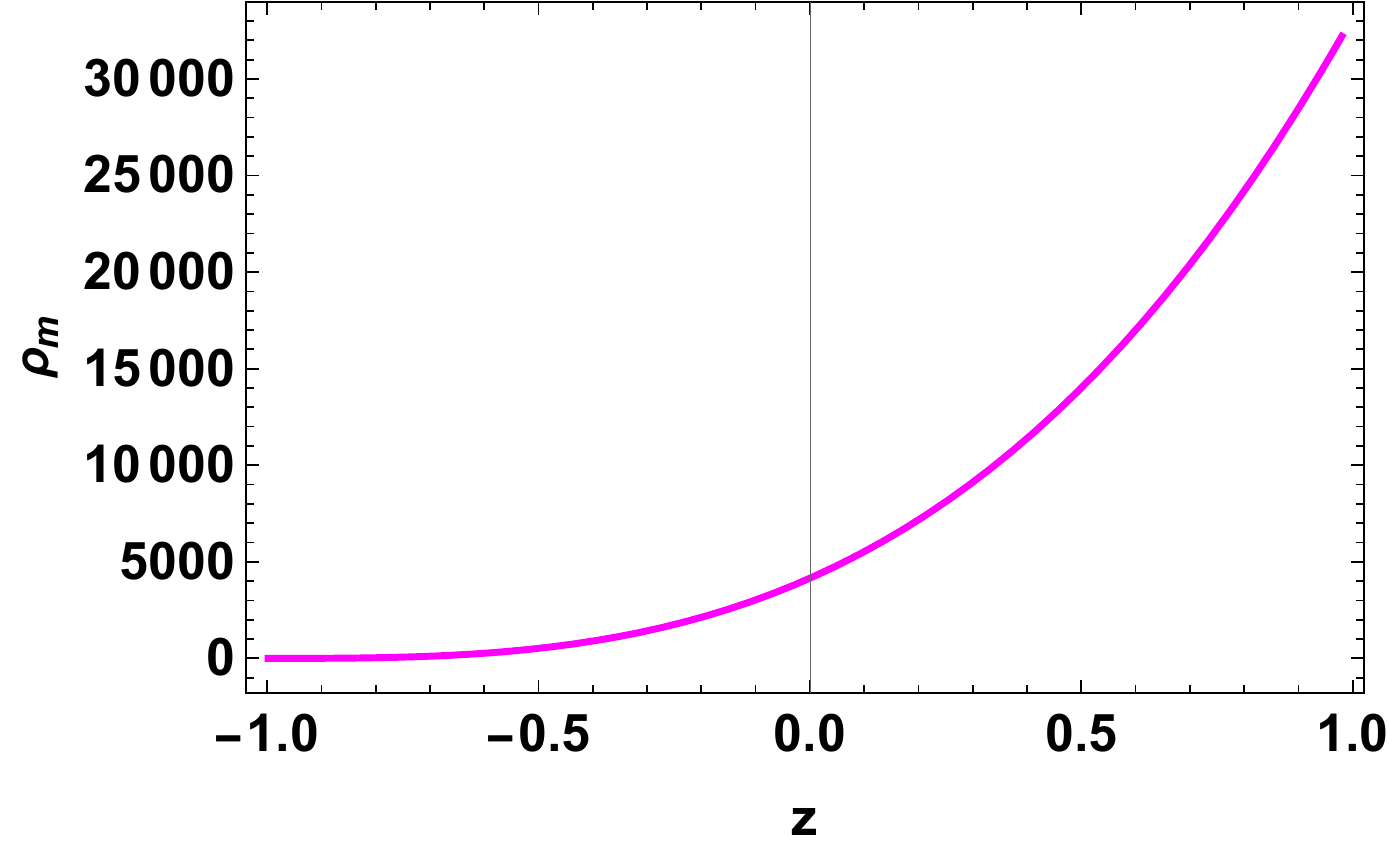}
\caption{Evolution profile of the cosmic energy density of matter vs redshift z.}\label{f10}
\end{figure}

\begin{figure}[H]
\includegraphics[scale=0.63]{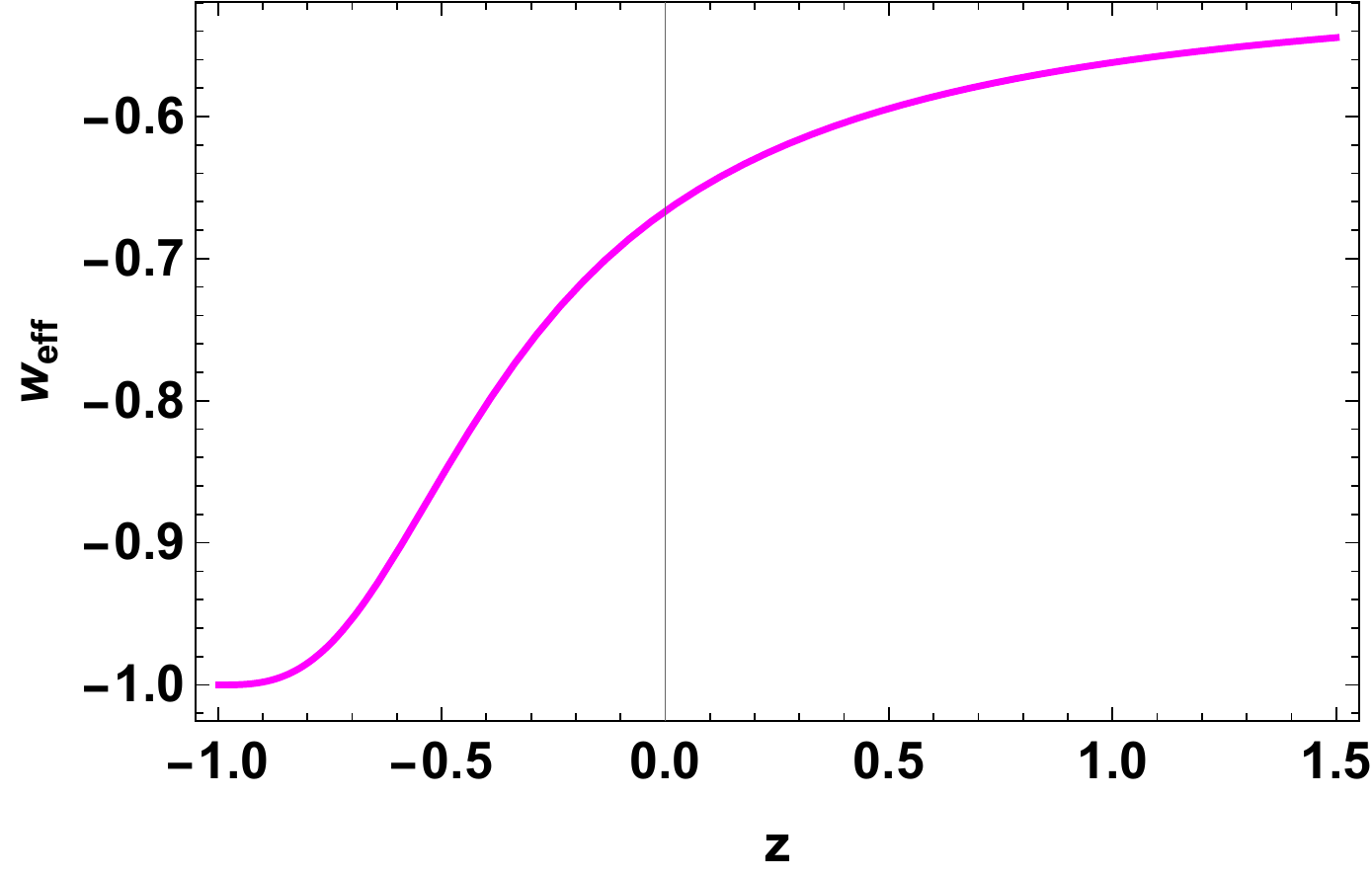}
\caption{Evolution profile of the effective EoS parameter vs redshift z.}\label{f11}
\end{figure}

\begin{figure}[H]
\includegraphics[scale=0.66]{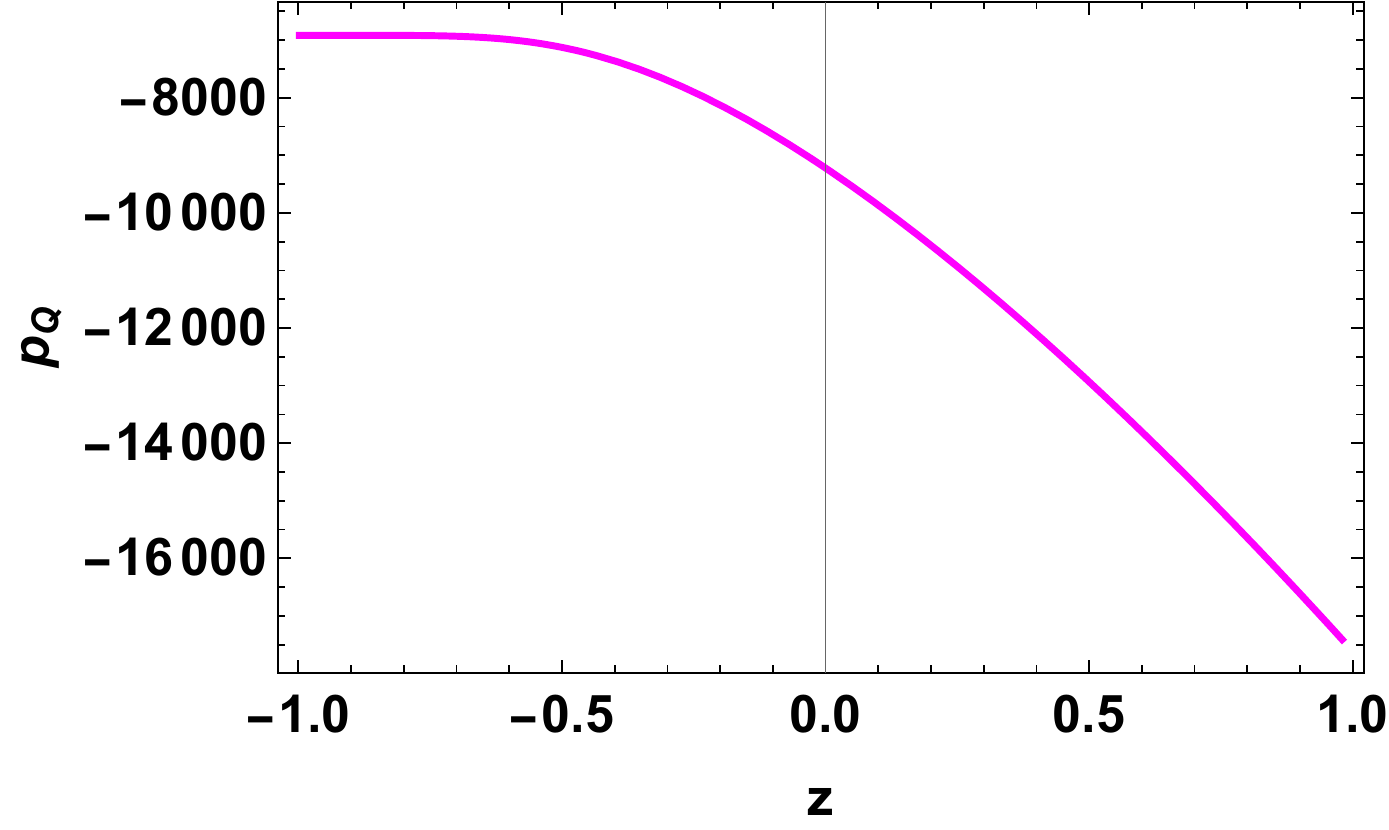}
\caption{Evolution profile of the cosmic pressure due to non-metricity vs redshift z.}\label{f12}
\end{figure}

\begin{figure}[H]
\includegraphics[scale=0.63]{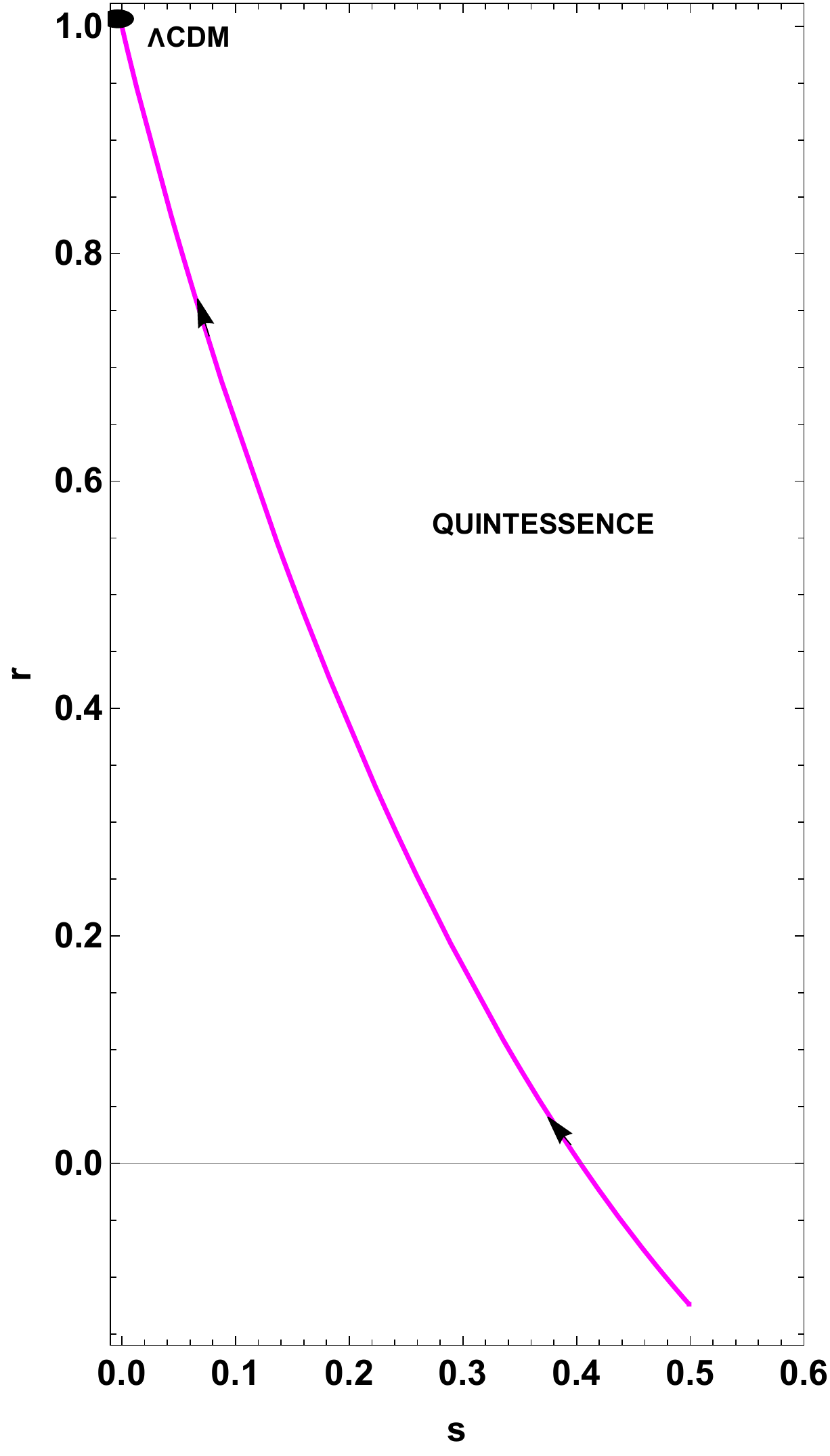}
\caption{Profile of the evolution trajectories of the given cosmological model in the $r-s$ plane . }\label{f13}
\end{figure}

From Fig \ref{f10} it is clear that the matter energy density of the universe decreases and vanishes in the far future with expansion of the universe while Fig \ref{f8} show that if the dark energy component is represented by the non-metricity then it will overcome all the energy content of the universe. Further, Fig \ref{f12} shows that the non-metricity component of the universe exhibit high negative pressure and this negative pressure decreases with the expansion of the universe. The plot for deceleration parameter in Fig \ref{f9} shows negative behavior that represents the accelerating expansion phase of the universe. However this $f(Q)$ model do not show the recent transition from early deceleration to present acceleration of the universe. Again, from Fig \ref{f11} it is clear that the present universe is in the accelerating phase. 
Lastly, Fig \ref{f13} represents the evolution trajectories of the given polynomial $f(Q)$ model that lies in the quintessence region $(r<1, s>0)$ and it will pass through the $\Lambda$CDM fixed point in the far future.

\section{Om Diagnostics}\label{sec5}

The Om diagnostic is another effective tool to differentiate cosmological models of dark energy \cite{Om}. It is simplest diagnostic as compared to statefinder diagnostic since it uses only Hubble parameter which requires first order derivative of cosmic scale factor. For spatially falt universe, it is defined as

\begin{equation}
Om(z)= \frac{E(z)^2-1}{(1+z)^3-1}
\end{equation}

\begin{figure}[H]
\includegraphics[scale=0.6]{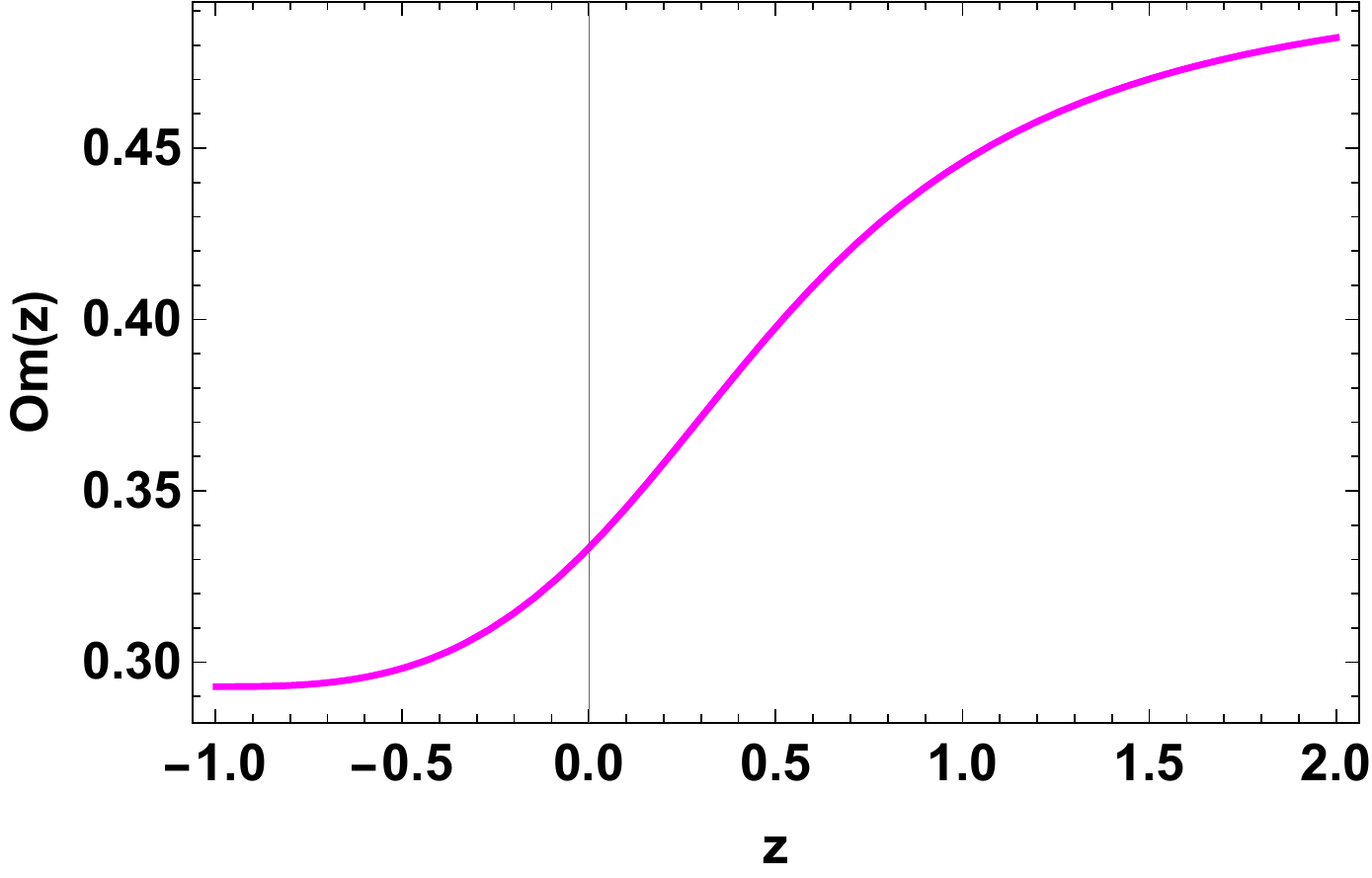}
\caption{Profile of the Om diagnostic parameter for the Cosmological Model I. }\label{f7}
\end{figure}

\begin{figure}[H]
\includegraphics[scale=0.62]{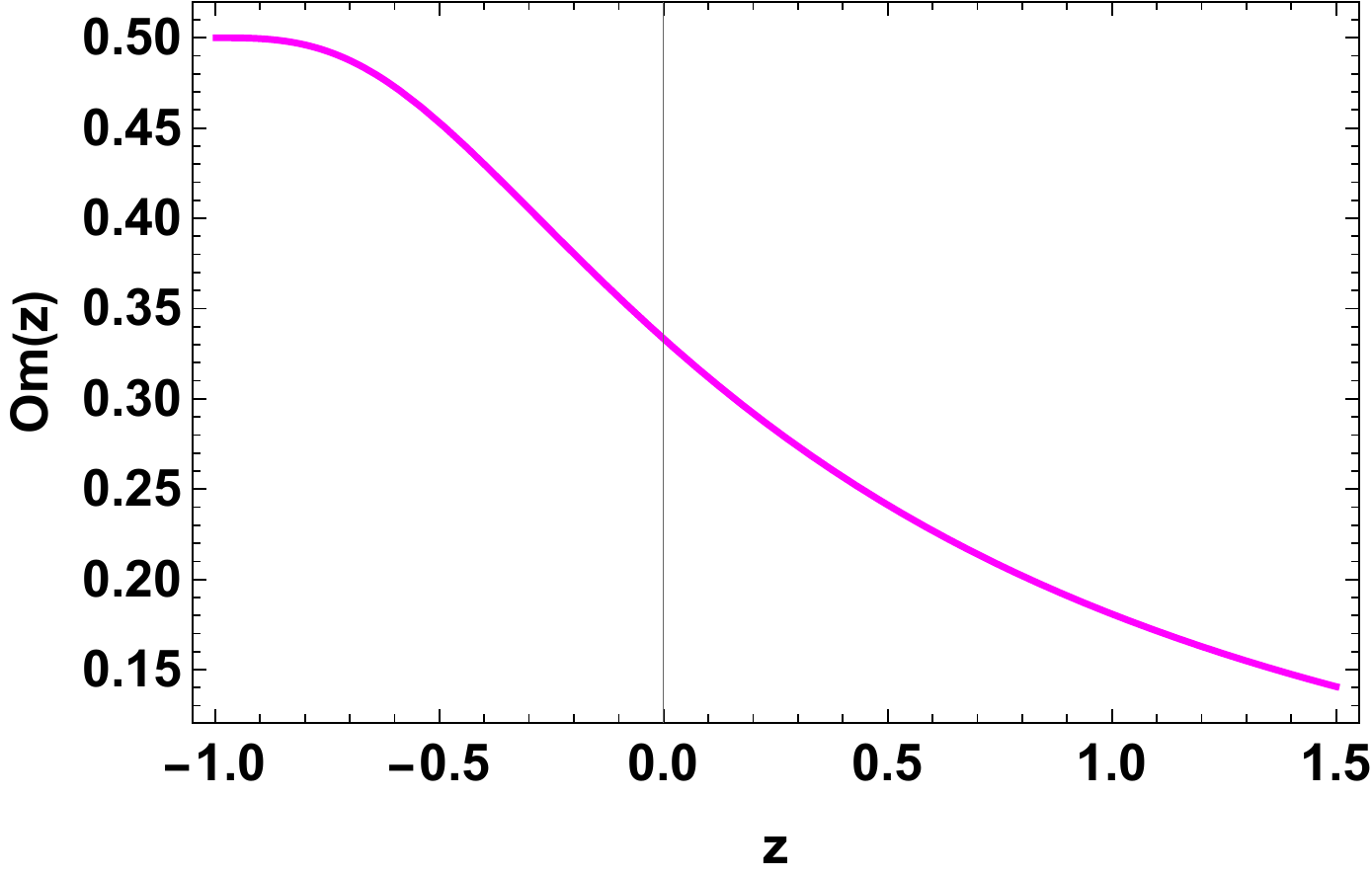}
\caption{Profile of the Om diagnostic parameter for the Cosmological Model II. }\label{f14}
\end{figure}

Here $E(z)=\frac{H(z)}{H_0}$ and $H_0$ is the Hubble constant. The negative slope of $Om(z)$ correspond to quintessence behavior whereas positive slope corresponds to phantom type behavior. The constant behavior of $Om(z)$ indicates the $\Lambda$CDM model. From Fig \eqref{f7} and \eqref{f14} we observed that the Om diagnostic parameter corresponding to the model I have positive slope while corresponding to model II have negative slope on the entire domain. Thus from Om diagnostic we can  conclude that the model I represents phantom type behavior while model II follows quintessence scenario.

\section{Conclusion}\label{sec6}

Cosmology has been struggling mainly for two issues: dark matter and dark energy. The dark matter does not interact electromagnetically while its gravitational effects can be detected. Still, we have not yet found dark matter as a particle of standard model \cite{CD,DS}. Modified gravity theory has been used also to explain dark matter effects. Further, it is a highly counter instinctive fact that our universe is accelerating. Although, cosmological constant in GR can well describe this dynamical effect the aforementioned issues related to $\Lambda$ motivate the search for an alternative explanation of the dark energy. As a consequence, so many dark energy models started appearing. The discrimination between the various dark energy models becomes necessary. Statefinder diagnostic is a convenient method that can differentiate between the various dark energy models.

In this article, we have performed a complete analysis of statefinder parameters for $f(Q)$ cosmology. We considered two $f(Q)$ models which contains a linear and a non-linear term of non-metricity scalar, specifically, $f(Q)=\alpha Q + \frac{\beta}{Q}$ and $f(Q)=\alpha Q + \beta Q^2$, where $\alpha$ and $\beta$ are free parameters and then we performed a complete diagnosis of the statefinder parameters.  We have plotted the evolution trajectory of our model in the $r-s$ plane and analyzed the behavior of different cosmological parameters. For both the models (see fig \ref{f1},\ref{f3}, \ref{f8}, and \ref{f10} ), we found that the matter energy density falls off as the universe expands while the dark energy due to non-metricity will overcome all the energy content of the universe. From Fig \ref{f4} and \ref{f11}, we found that the present universe is in the accelerating phase. In addition, Fig \ref{f2} show that our universe has experienced a transition from decelerated to accelerated phase in the recent past while Fig \ref{f9} do not show the transition phase. Also, From Fig \ref{f5} and \ref{f12} we observed that the non-metricity component of the universe exhibit high negative pressure. Further, Fig \ref{f6} represents the evolution trajectories of the model I which lies in the chaplygin gas region $(r>1, s<0)$ and it will pass through the $\Lambda$CDM fixed point in the far future whereas the evolution trajectories of the model II in Fig \ref{f13} lies in the quintessence region $(r<1, s>0)$. Finally, from Om diagnostic (see fig \ref{f7} and \ref{f14}) we can  conclude that the model I represents phantom type behavior while model II follows quintessence scenario. The cosmological $f(Q)$ models presented in this work have great significance. A power law correction to the STEGR will give rise to branches of solution applicable either to the early universe or to late-time cosmic acceleration. Our cosmological model I provides a correction to the late-time cosmology, where they can give rise to the dark energy, whereas model II is relevant to the early universe with potential applications to inflationary solutions \cite{Lavi}. Furthermore, another important aspect of our considered $f(Q)$ models is that
throughout the evolution of the universe, the varying growth index of matter perturbations is smaller than that of the $\Lambda$CDM for our cosmological model I, whereas larger for model II, which can be helpful to estimate the distribution of matter in the universe \cite{Dutta}.

\section{Data Availability Statement}
There are no new data associated with this article.

\section*{Acknowledgments} \label{sec7}
RS acknowledges University Grants Commission(UGC), New Delhi, India for awarding Junior Research Fellowship (UGC-Ref. No.: 191620096030). We are very much grateful to the honorable referees and to the editor for the illuminating suggestions that have significantly improved our work in terms of research quality, and presentation.

%
%
%

\end{document}